# Threads free falling as a macroscopic replica of random walk phenomena and crystallization process in one dimensional space

H.D. Rahmayanti, R. Munir, E. Sustini, and M. Abdullah[a]

Department of Physics, Institut Teknologi Bandung, Jl. Ganesa 10, Bandung 40132, Indonesia

[a]Email: mikrajuddin@gmail.com

## ABSTRACT

We observed the conformation of threads that free released from different altitudes. We also explore the effect of wetting liquid on the conformation developed. When released from an altitude, the thread conformation changes when free falling and reaches the stable conformation (identified by a constant end-to-end distance) after fall over a period of time. We identified that the conformation of the threads replicated the conformation of long polymer chains. Therefore, we applied a two-dimensional random walk model to explain the stable conformation where the thread length is treated as the random walk time (number of polymer segment). A two-dimensional random walk model was used since during free fall, the conformation is assumed to develop on the horizontal plane. Surprisingly, we get the scaling power that is exactly similar to the scale power of two dimensional random (walk). By fitting how end-to-end distance changes with time, we obtain an equation that is exactly similar to modified Avrami equation for process of phase transformation in one-dimension space ($D = 1$), where the exponential power is $n = D+1 = 2$. Therefore, we can state that the time evolution of thread conformation process is identical to crystallization process in one dimensional space. This is very interesting finding, showing that the microscopic process (crystallization) can be replicated in macroscopic scale.

# INTRODUCTION

If we leave a thread that was initially stretched horizontally to fall freely, we will observe that the shape will change to become curly. Specifically, the end to end distance becomes shorter and the shape will be likely the shape of long polymer chains observed with the electron microscope. Interestingly, the shape shrinks as time increases and likely to reach the stable size, where further increasing in time does not change the size, but the curly orientation might change. It means that, the threads will evolve into a final size.

The shapes of polymer chains have been frequently described using random walk model [1,2]. In many situations such model was successfully describing the chain conformation, including an accurate estimation of the end to end distance and radius of gyration as function of chain length [3,4]. Since the shape of thread left to free fall mimics the shape of polymer chain, it is then challenging to ask the potential application of random walk model to explain the conformation of polymer chain after reaching the stable size (after left to fall from high altitude). This exploration is interesting since it becomes a simple macroscopic demonstration of random walk model that can be observed by naked eye.

As we have mentioned many times, indeed, there are many common phenomena around us that might become an interesting topic of research which in many cases have been unconsidered by many people [5-12]. Leaving the threads to free fall from an altitude might be unimportant phenomenon, since so many people have experienced such process and neither interesting thing can be extracted. However, here we will show that this phenomenon may attract a lot of attention since it also shows many interesting physical phenomena. Similar explorations of "common phenomena" to show rich of physics have been reported such as walking with coffee [13], capillary force repels coffee-ring effect [14], fingering inside the coffee ring [15], shapes of a suspended curly hair [16], Bending of sparklers [10], wringing of wet cloth [7].

This phenomenon thread free fall can be investigated further by asking what is the effect of the thread materials and the level of wetting applied to the threads (since different wetting levels lead to different elastic properties of the thread, including the effect of different wetting materials) on the final conformation of the threads. Is the behavior approximate the behavior of random walk phenomena in microscopic scale that have been deeply explored by many scientists within a century?

The purpose of this work is to explore the shape of thread left to free fall and to find its correlation with the random walk phenomena. This work also proved the random walk phenomena in macroscopic scale. We performed simple experimental and derived mathematical formulae to describe the observed data.

**EXPERIMENTAL**

In this experiment, we used three types of threads: wool, mélange and multifilament. The length of the threads is varied: 0.5 m, 0.75 m and 1 m. We applied three mechanical conditions to the threads: dry, wetted with water, and wetted with alcohol. The thread released from up to altitude of 22.5 meters and positioned horizontally. The time for the thread to touch the ground is also recorded. After the threads touched the ground, the end to end distance is measured. Several repetitions were conducted for each experimental condition.

We also took the pictures of threads after touching the ground when fell at different altitudes to show the conformation. The threads were initially immersed in a luminescent liquid and the pictures were taken in the dark under specific light illumination to show the conformation of the threads.

**EXPERIMENTAL RESULT**

Figure 1 shows the images of different threads: (a) wool, (b) mélange and (c) multifilament that was released at different altitudes: (i) 0.5 m, (ii) 1.5 m and (iii) 2.5 m. The pictures were taken in the dark and using a specific light source to glow three threads. The three threads had the same length (1 m), and the images were recorded after the threads touched the ground. Clearly shown that all threads have formed a conformation of nearly similar to polymer chain conformation in stable state , especially for the threads that have been released from high altitudes. Generally, the curling conformation increases when the altitude increases. This means that the initial state when the threads form straight lines is very unstable states. Under freely to release (assuming the interaction of the threads with molecules in air is negligible), the threads evolved into stable conformation when time evolved. We can also see that the conformation of the threads is closely related to the random walk conformation.

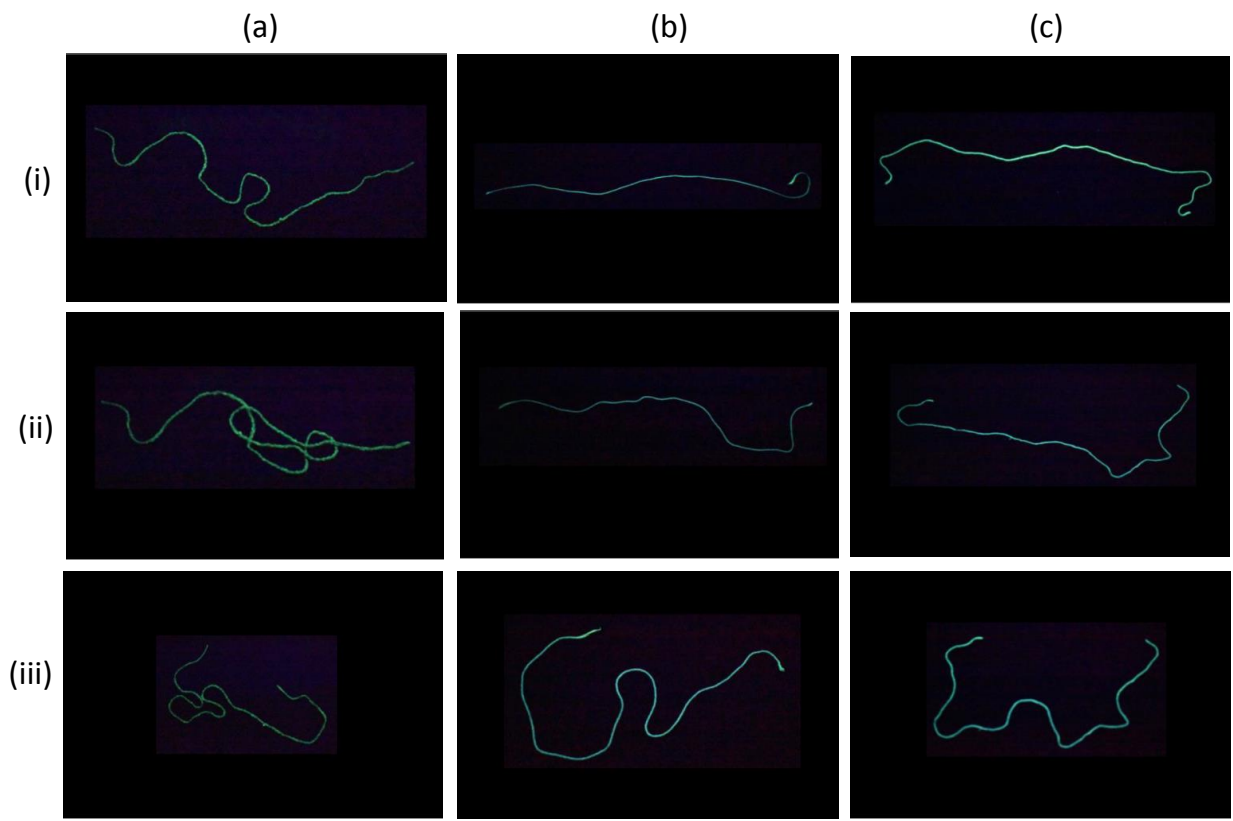

FIG. 1. The topologies of threads after touching the ground. Thre threads glow under specific light source. We used different threads: (a) wool, (b) mélange and (c) multifilament, released at different altitudes: (i) 0.5 m, (ii) 1.5 m, dan (iii) 2.5 m.

We also inspect the final conformation when the thread length is varied. Figure 2(left) shows the final topologies of multifilament thread released at an altitude of 1.0 meters. Three different lengths were used: 0.5 m, 0.75 m, and 1.0 m. Longer threads correspond to random walk at longer time (time is proportional to the thred length). Thus from the figure (a) to (c) we may state that a random walk at increasing time steps (time increases from (a) to (c)).

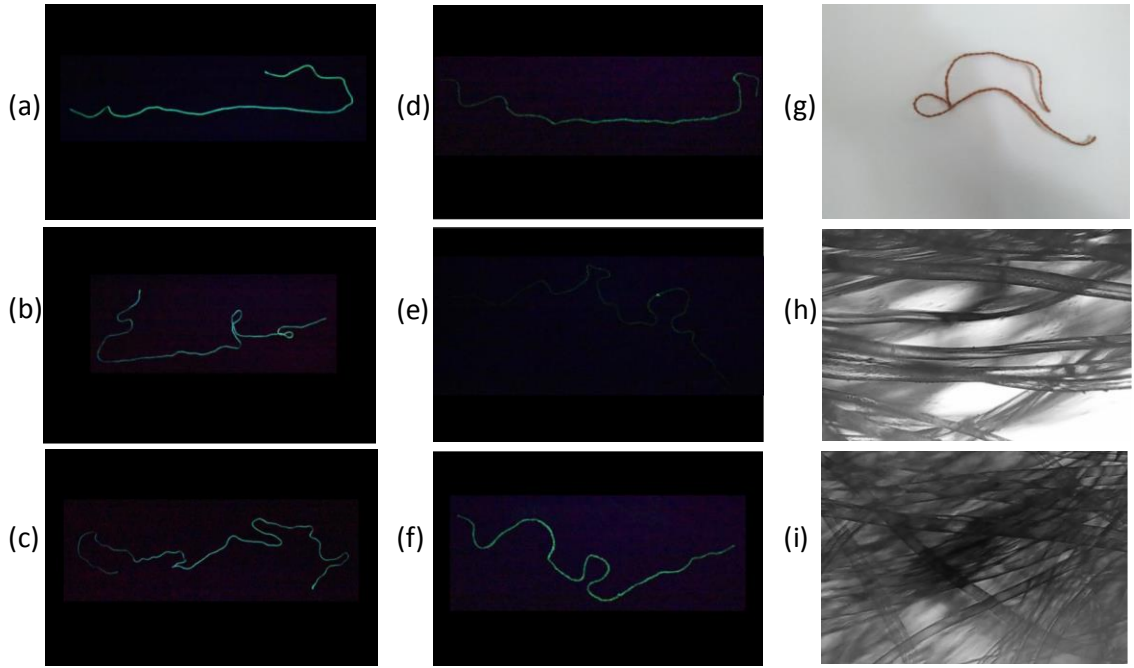

FIG. 2. (left) Images of multifilament thread at different length: (a) 0.5 m, (b) 0.75 m and (c) 1.0 m, released at the same altitude of 1.0 m. (middle) Images of wool of 0.5 m lengths of different wetting liquid released at an altitude of 0.5m: (d) dry wool, (e) wool wetted with alcohol, (f) wool wetted with water. (right) Images of (g) dry wool recorded using digital camera, images of (h) wool wetted with alcohol, and (i) wool wetted with water both recorded using an optical microscope.

The effect of the wetting liquid on the conformation was also inspected. We compared the topologies of dry wool, wool wetted with alcohol, and wool wetted with water. Three threads of the same length 0.5 m were released from the same altitude of 0.5 m. As shown in Fig. 2 (middle), dry wool tends to form a straight conformation, while wet wool with water tends to form a more curly conformation. It means that the dry wool is much harder to be bent, and the wool wetted with water is the most easily bent.

We also recorded the microscopic images of the sample. At present we show the images of wool as shown in Fig. 2(right): (a) camera picture of dry wool, (b) microscope image of wool wetted with alcohol, and (c) microscope image of wool wetted with water. It is likely that the wool absorbs more water rather than alcohol, causing the wool wetted with water becomes much more weaker than the dry wool or the wool wetted with alcohol.

We then measured quantitatively the distance between threads end at different conditions: different threads, different altitude for releasing, and different wetting liquid. This distance corresponds to end-to-end distance from the random walk to inspect weather the curly conformation of the threads has a correlation with the random walk. The most important parameter is the power factor relating the random walk time and the end-to-end distance. If we divide the threads into a large number of segments we can relate the segment length correspond to distance traveled at one-time step and the thread length corresponds to the total time of the random walk. In the random walk model, the end-to-end distance satisfies $\sqrt{\langle r^2 \rangle} \propto t^\kappa$ with t is time and κ is scaling factor around 0.5 for simple random walk. In the present work, we need to inspect is the similar relation will also been obtained, i.e. $\sqrt{\langle r^2 \rangle} \propto L^\kappa$ with $L$ is the thread length and $\kappa$ is the scaling parameter for two-dimensional random walks. We must understand that the conformation of the released thread will patch the two-dimensional random walk due to gravitation attraction will attract the thread element simultaneously (under assumption the friction by air is identical to all segments).

We dropped the thread in horizontal straight position. The shape evolved to be curly and the end to end distance decreases continuously. The question is how long the time required by the thread to get the stable conformation, to mean the end to end distance no longer changes. For this purpose, we dropped the threads from a storey building up to an altitude of 22.5 meters from the ground. Figure 3 shows the data of end to end distance between threads from different materials in three length variations and the same thread wetted with different liquids. From all figures, it becomes clear that the stable end to end distances have been obtained after the threads fall about 5 m. The final end to end distance became relatively constant after dropping from an attitude 5 m to 22.5 m in this experiment. It is also clear that the end to end distance decreases as the length of the thread decreases to confirm the relation $\sqrt{\langle r^2 \rangle} \propto L^\kappa$ seems to be acceptable. What we need to explore is what is the value of parameter $\kappa$. Is this parameter consistent with the corresponding parameter belongs to the random walk process?

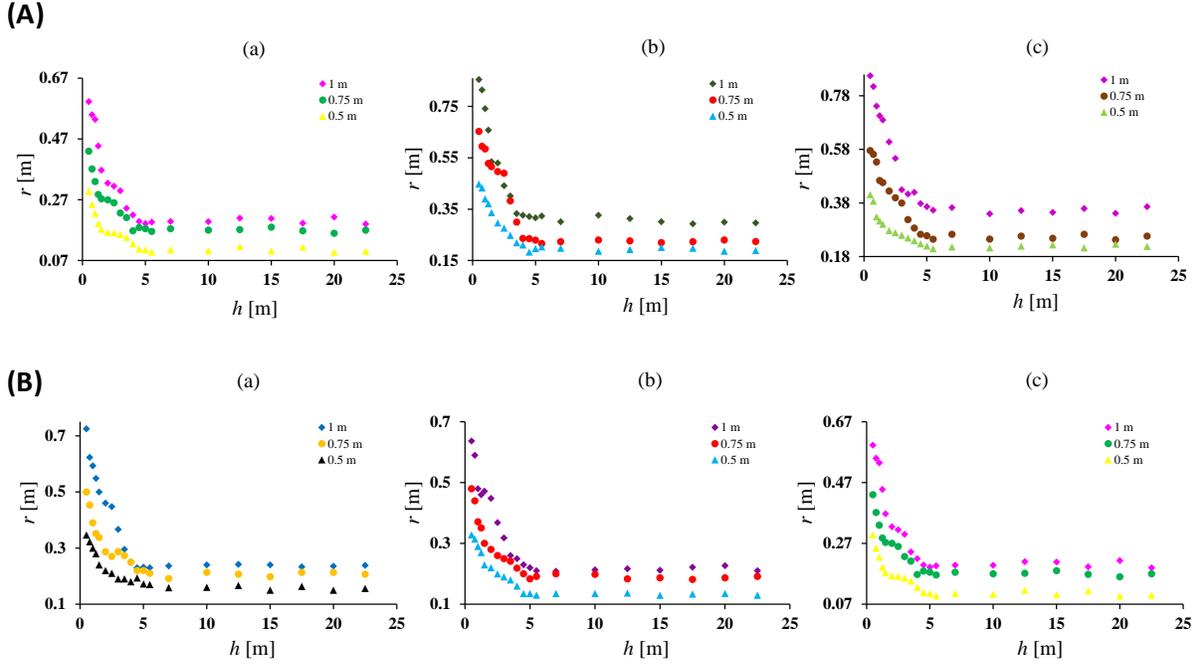

FIG. 3. Effect of altitude on the end to end distance of thread. Upper figures (A): (a) wool thread, (b) mélange thread and (c) multifilament thread. All threads are varied from three length variations (0.5 m, 0.75 m and 1.0 m). Bottom figure (B): dry wool, (b) wool wetted with alcohol and (c) wool wetted with water. Symbols are measured data while curves are fitting results.

At the same time we also measured the relationship between the falling time and and falling distance. For example, Fig. 4 shows the plot of falling time and square of falling distance. As comparison, for a free fall (when neglecting the air resistance) at zero initial speed we have a simple relation $h = gt^2/2$ at $t = \sqrt{2/g}\, h^{1/2}$ with $h$ is falling distance, $t$ is falling time and $g$ is the acceleration of gravitation. It is clear that for this condition, $t$ is a liner function of $h^{1/2}$, indicating that the air resistance can not be ignored. The presence of air resistance is indicated by the rate increase in $h^{1/2}$ decreases as time increases.

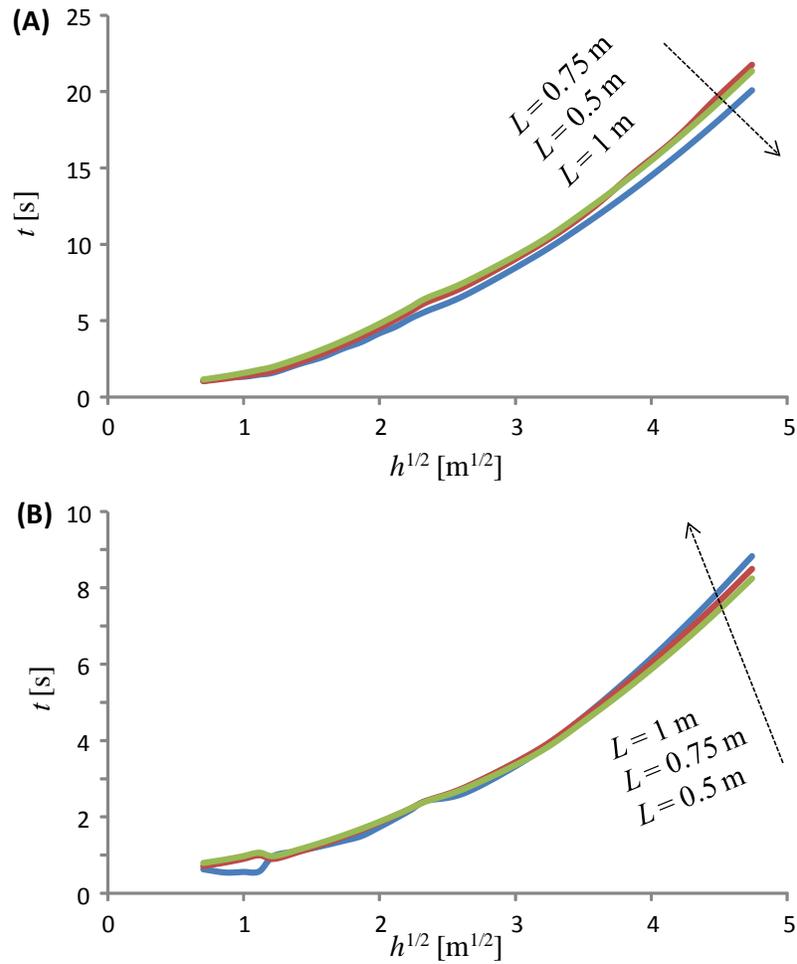

FIG. 4. Plot of falling time and square of falling distance of: (A) dry wool and (B) wool wetted with alcohol.

Figure 5 shows the plot of $\ln r$ ($r$ is the average end to end distance) with respect to $\ln L$ for (a) wool thread wetted with water, (b) mélange thread wetted with water, and (c) multifilament thread wetted with water. Symbols are the measurement results. It is clear that the data strongly show the linear relationship between $\ln r$ and $\ln L$. Based on the fitting results, we obtained the gradients of the lines are $\kappa = 0.536$, 0.725, and 0.656 for wool thread wetted with water, mélange thread wetted with water, and multifilament thread wetted with water, respectively. Three collected data satisfy the fitting equations: $\ln r = 0.536 \ln L - 1.668$, $\ln r = 0.725 \ln L - 1.185$, and

$\ln r = 0.656\ln L - 1.053$, for wool thread wetted with water, mélange thread wetted with water, and multifilament thread wetted with water, respectively.

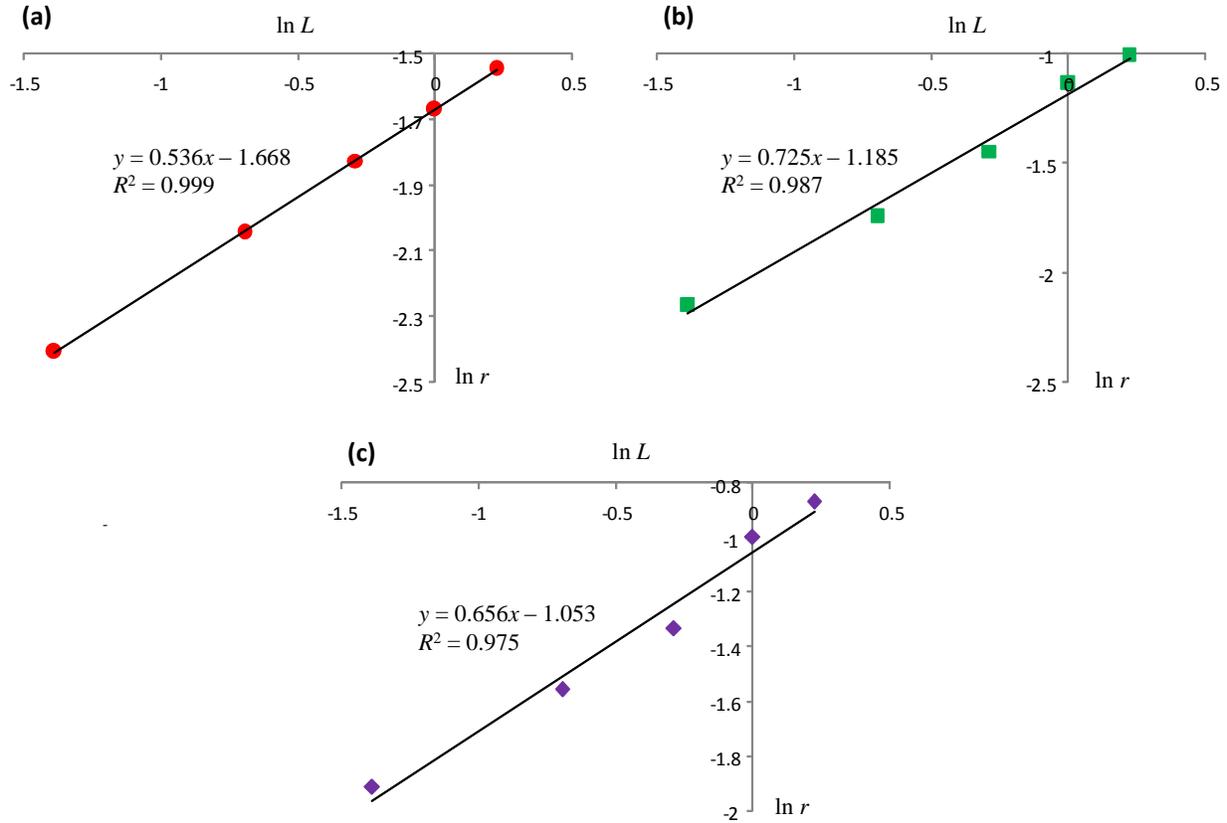

FIG. 5. Effect of thread length of the end to end distance by assuming that the stable distance has been achieved: (a) wool thread wetted with water, (b) mélange thread wetted with water, and (c) multifilament thread wetted with water. We obtained the fitting equations $\ln r = 0.536\ln L - 1.668$, $\ln r = 0.725\ln L - 1.185$ and $\ln r = 0.656\ln L - 1.053$, for wool thread wetted with water, mélange thread wetted with water, and multifilament thread wetted with water, respectively

Figure 6 is the similar data for wool wetted with different liquid: (a) dry wool, (b) wool wetted with alcohol, and (c) wool wetted with water. It is also clear that the data strongly show the linear relationship between $\ln r$ and $\ln L$. Three collected data satisfy the fitting equations:

$\ln r = 0.509 \ln L - 1.311$, $\ln r = 0.547 \ln L - 1.597$, and $\ln r = 0.536 \ln L - 1.669$, for dry wool, wool wetted with alcohol, and wool wetted with water, respectively

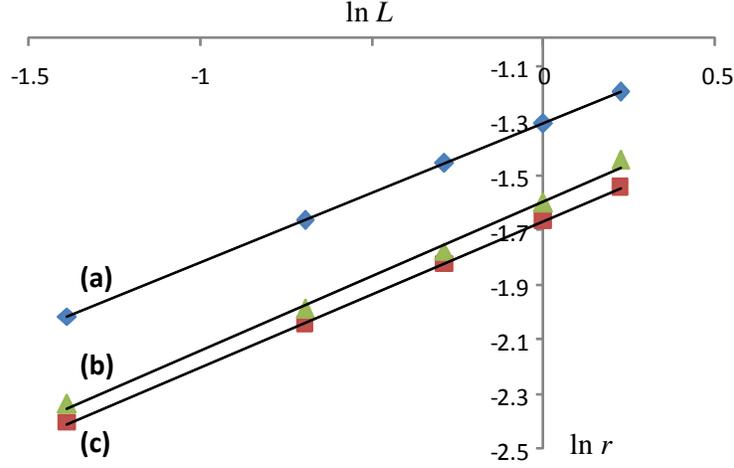

FIG. 6. Effect of thread length on the end to end distance for (a) dry wool, (b) wool wetted with alcohol, and (c) wool wetted with water. We obtained the corresponding fitting equations: (a) $\ln r = 0.509 \ln L - 1.311$, $R^2 = 1.0$; (b) $\ln r = 0.547 \ln L - 1.597$, $R^2 = 0.995$; and (c) $\ln r = 0.536 \ln L - 1.669$, $R^2 = 0.999$.

## MODELLING AND DISCUSSION

Since we have suspected that the thread conformation will approach the conformation of two dimensional random walks, it is then interesting to compare our results of the random walk model. The repulsive energy for a walk of $N$ steps stretching out a distance $r$ in a space with fractal dimension $\bar{d}$ was approximated as [17]

$$U \approx N^2 r^{-\bar{d}} \qquad (1)$$

The entropy was estimated at the assumption of a Gaussian distribution of the distance $r$ traveled by a random walk after $N$ steps as

$$S \approx -r^2 N^{-2/d_w} \qquad (2)$$

with $d_w$ is the diffusion dimension, defined by $r^{d_w} \approx t$ for very large $t$ (time elapsed for random walk). Both $\bar{d}$ and $d_w$ are related by equation $\bar{d} + \zeta = d_w$, with $\zeta$ is the resistivity exponent defined by $\Omega(r) \approx r^{\zeta}$, for the resistivity $\Omega$ between two points at a distance $r$. Based on Eq. (1) and (2) we estimated the Helmholtz free energy $F = U - TS$ as

$$F \approx N^2 r^{-\bar{d}} + Tr^2 N^{-2/d_w} \qquad (3)$$

with $T$ is temperature. The equilibrium states is obtained by minimizing $F$ to produce $-\bar{d}N^2 r^{-\bar{d}} + 2TrN^{-2/d_w} \approx 0$, and resulting

$$r \approx \left(\frac{\bar{d}}{2T}\right)^{1/(2+\bar{d})} N^{\kappa} \qquad (4)$$

with

$$\kappa = \frac{2(1+1/d_w)}{2+\bar{d}} \qquad (5)$$

Eqs. (1) to (5) have been derived from assumption of free random walk. However, for threads used at present work, the energy depends on the thread properties such as elasticity (although the modulus of elasticity might be very small especially for the wet threads) and wetting level. Dry and wet threads should have different repulsive energy for the same stretching distance. Therefore, for threads we propose the repulsive energy is rescaled as following

$$U \approx \phi N^2 r^{-\bar{d}} \qquad (6)$$

To give the Helmholtz free energy as

$$F \approx \phi N^2 r^{-\bar{d}} + Tr^2 N^{-2/d_w} \qquad (7)$$

with $\phi$ is a parameter accounting for thread properties as mentioned above. Using the same procedures as for obtaining Eq. (4), the distance between thread ends is estimated as

$$r(\phi, N) \approx \left(\frac{\phi\bar{d}}{2T}\right)^{1/(2+\bar{d})} N^{\kappa} \qquad (8)$$

If the thread is considered as connected of N identical segments of the same length, a, we can write $N = L/a$ with L is the thread length. Substituting into Eq. (8) we have

$$r(\phi, L) \approx \psi \phi^{1/(2+\bar{d})} L^{\kappa} \qquad (9)$$

with

$$\psi \approx \frac{1}{a^{\kappa}} \left( \frac{\bar{d}}{2T} \right)^{1/(2+\bar{d})} \qquad (10)$$

Wet threads are much easier to be bent rather than dry threads. The parameter $\psi$ is a material constant. Thus $\phi$ decreases as the wetting level increases. From Eq. (9) we can write

$$\ln r(\phi, L) \approx \ln\left[\psi \phi^{1/(2+\bar{d})}\right] + \kappa \ln L \qquad (11)$$

Dekeyser et al [17] have shown that, for two-dimensional random walk, the following estimated values are acceptable: $d_w \approx 3$ and $\bar{d} \approx 2$. Substituting into Eq. (5) one has

$$\kappa \approx \frac{2(1+1/3)}{2+2} = 0.666 \qquad (12)$$

Surprisingly, this value is very close to the fitting result of data onto Fig. 5 between 0.536 – 0.725. For two-dimensional random walk, it is also accepted $\bar{d} = 4/3$ [18,19] and $d_w \approx 2.8$. Using these values, we have other estimation for $\kappa \approx 0.8$ which is also closed to the fitting result.

It is clear from Fig. 6 that $d_w$ is nearly independent of the wetting level, indicated by the gradients for three conditions changes only slightly (0.509, 0.547, and 0.536). The wetting level only affects the end to end distance.

Let us consider the effect of wetting on the energy of the thread. Based on Fig. 5 and Fig. 6 we clearly obtained a linear fitting of $y = ax + b$ for all data, where $y = \ln r(\phi, L)$, $x = \ln L$, $a = \kappa$ and $b \approx \ln\left[\psi \phi^{1/(2+\bar{d})}\right]$. The last relationship can be rewritten as $\psi \phi^{1/(2+\bar{d})} \approx \exp(b)$. Substituting Eq.

(10) into Eq. (11) we can write $\phi^{1/(2+\bar{d})} \propto a^{\kappa} \exp(b)$. By assuming that $a^{\kappa}$ is nearly the same for the same threads at all conditions (as shown in Fig. 6 where $\kappa$ is nearly the same for dry, alcohol wetted and water wetted) we can then approximate $\phi^{1/(2+\bar{d})} \propto \exp(b)$, or

$$\phi \propto e^{(2+\bar{d})b} \tag{13}$$

Based on the fitting results as displayed in Fig. 6, and using $\bar{d} \approx 2$, wetting the wool with alcohol ($b = -1.597$) reduces the energy of the thread compared to the dry wool ($b = -1.311$) by 3.14 and wetting the wool with water ($b = -1.669$) reduces the energy of the thread compared to the dry wool by 4.19.

In experiment, we observed the distance between threads ends evolved with time. As falling distance increases, the end-to-end distance decreases to approximate a certain value for the very long time. Since the falling distance increases with time, we then conclude that the end-to-end distance evolves with time. It is then challenging to explore any equation that can describe the evolution of the thread ends distance.

The reduction of thread ends distance can be compared with the process of shrinkage. Thus, the equation describing the rate of shrinkage might be adopted. Weir proposed an equation to explain the shrinkage of tendon collagen and proposed a simple equation to explain the length of the tendon as [20] $\lambda = (\lambda_0 - \lambda_\infty)\exp(-Kt) + \lambda_\infty$, with $\lambda_0$ is the initial length of the tendon and $\lambda_\infty$ is the final length of the tendon (when $t \to \infty$) and $K$ is the rate constant. We assumed the same equation applies for end to end distance as

$$r(L,t) = (L - r(\phi, L))\exp(-Kt) + r(\phi, L) \tag{14}$$

with $K$ in general is a function of $t$.

To confirm Eq. (14) we measured the time required by the threads to fall at different distances. Figure 7 shows the plot of end-to-end distance of wool wetted with water and multifilament wetted with water as function of falling time. Each material is varied from three different lengths. Symbols are measurement results and curves are fitting curve using Eq. (14). It is obvious that all data can be well fitted with Eq. (14) that producing $R^2 > 0.87$ (mostly $> 0.95$).

Let us inspect Fig. 7(A). For $L = 0.5$ m, 0.75 m, and 1 m we have $r(\phi,0.5) = 0.12$ m, $r(\phi,0.75) = 0.17$ m, and $r(\phi,1) = 0.2$ m, respectively. If we assume the relation of $r(\phi, L) \propto L^\kappa$ is satisfies with $\kappa$ has been aproximated to be around 0.666 then the we must have $r(\phi, L_1)/r(\phi, L_2) \propto (L_1/L_2)^{0.666}$. Using $L_1 = 0.75$ m and $L_2 = 1$ m we have $r_1/r_2 = 0.85$ while $(L_1/L_2)^{0.666} = 0.83$. Then using $L_1 = 0.5$ ans $L_2 = 1$ we have $r_1/r_2 = 0.6$ while $(L_1/L_2)^{0.666} = 0.63$. These results shown a very nice consistence. Another example is Fig. 7(B). For $L = 0.5$ m, 0.75 m, and 1 m we have $r(\phi,0.5) = 0.21$ m, $r(\phi,0.75) = 0.254$ m, and $r(\phi,1) = 0.31$ m, respectively. Using $L_1 = 0.75$ m and $L_2 = 1$ m we have $r_1/r_2 = 0.819$ while $(L_1/L_2)^{0.666} = 0.83$. Then using $L_1 = 0.5$ ans $L_2 = 1$ we have $r_1/r_2 = 0.677$ while $(L_1/L_2)^{0.666} = 0.63$. These results also shown a very nice consistence.

Instead of using constant $K$, in this model we identified that the better fittings have been obtained using $K$ of a linear function of time, $K = pt + q$ with $p$ and $q$ are constants. Table 1 is the list of parameters used to get better fitting for data in Fig. 7. In all cases we obtained that $K$ increases in time. We observed, the values of each parameter in all thread lengths are nearly similar.

Since $p > 0$ and $q < 0$ for all fitting conditions we will identify that if $t < -q/2p$, the end-to-end distance increase with time since leaving the threads to free fall. This is impossible condition since the maximum end-to-end distance is equal to the thread length. Therefore, we may say that the fitting equations apply only for $t > -q/2p$. Furthermore, if we inspect the value of parameter $p$ and $q$ we will find that $-q/2p$ is very small. For example, the largest one is $1.0706/(2\times1.9758) = 0.27$ s. Other values are less than 0.15 s.

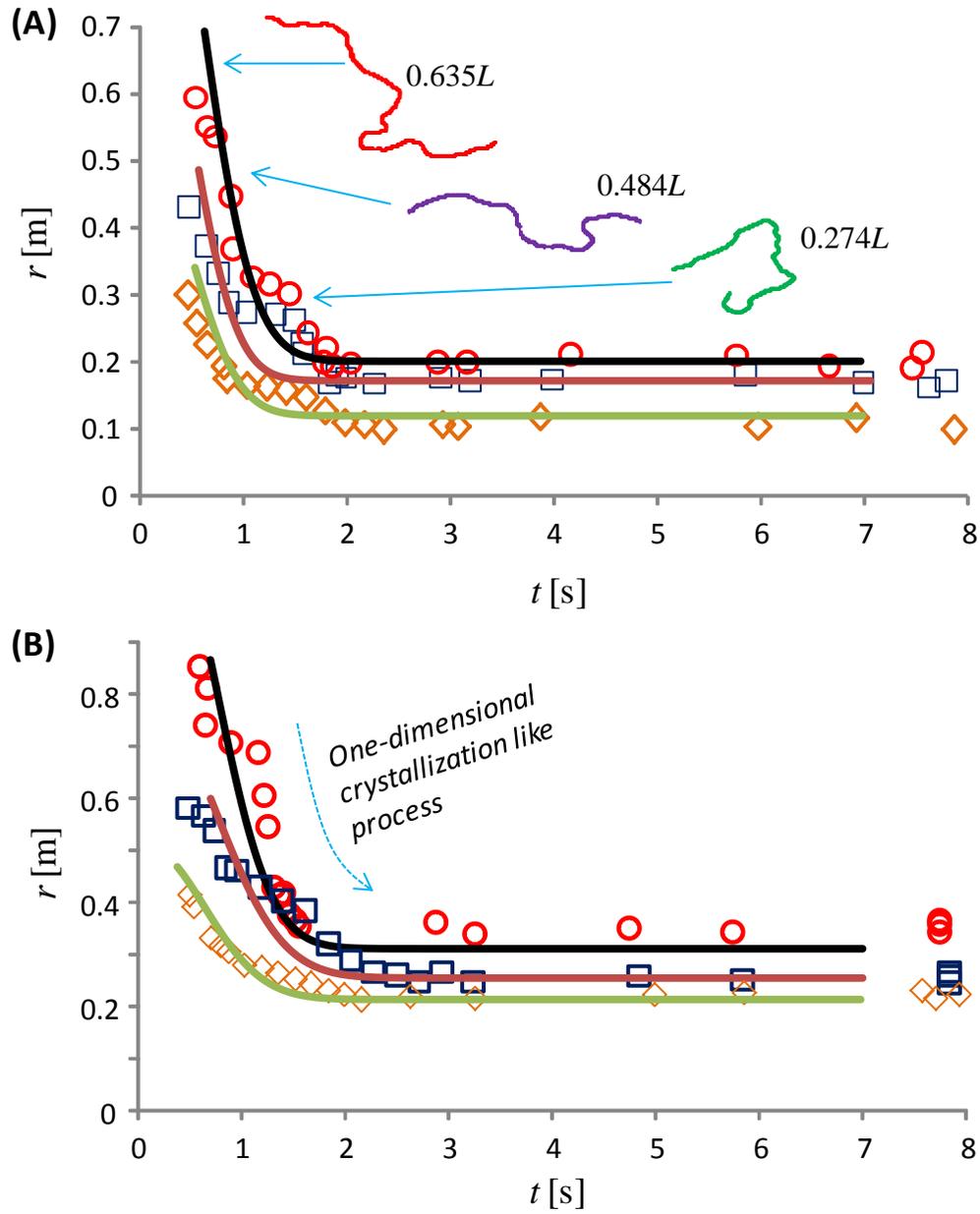

FIG. 7. Plot of end-to-end distance for: (A) wool wetted with water and (B) multifilament wetted with water as function of falling time. Each material is varied from three different lengths: (top) 1 m length, (middle) 0.75 m length, and (bottom) 0.5 m length. Symbols are measurement results and curves are fitting curve using Eq. (14). Inset in Fig. 7(A) is the simulated thread conformation at different end to end distance.

TABLE 1. Parameters used for fitting data in Fig. 7.

| Figure | Position | p | q |
|---|---|---|---|
| 7(A) | top | 2.1980 | -0.6838 |
|  | middle | 2.2888 | -0.6290 |
|  | bottom | 2.5552 | -0.415 |
| 7(B) | top | 1.9758 | -1.0706 |
|  | middle | 1.2983 | -0.3881 |
|  | bottom | 1.6011 | -0.3291 |

Based on the above discussing, we can strongly state that Eq. (14) has been able to explain the evolution of thread end-to-end distance with time. Indeed, some discrepancies still occur due to inaccuracy in experiment. The possible source of accurate was the measurement of time. The time was measured by two people, one placed at the living position (up to 22.5 meters above the ground) and one person stayed on the ground. During thread falling, the air resistance might also different size the continuous change in the thread conformation until touching the ground.

Let us further explore Eq. (14). Using the linear function of $K$ as results of fitting for all data, we obtain the following equation

$$Y = 1 - \exp\left(-p(t-t_m)^2\right) \tag{15}$$

with $t_m = -q/2p$ and

$$Y = 1 - \frac{r - r(\phi, L)}{L - r(\phi, L)} \exp\left(-\frac{q^2}{4p}\right) \tag{16}$$

Eq. (15) is exactly similar to the modified Avrami equation $Y = 1 - \exp(-p(t-t_m)^n)$ with $t_m$ is known as the incubation time [21] and $n = D+1$ with $D$ is the dimension of space in which crystallization occurs. The Avrami equation described the transformation of material from one phase to another phase at constant temperature, and $Y$ determines the fraction of new phase that has been formed at time $t$. Since we are dealing with thread having one dimensional behavior then $D = 1$ or $n = 2$, exactly the same as Eq. (15). Therefore, we can state that the process of thread conformation is likely a process of phase transformation in one-dimension space. The complete crystallization can be associated when the end-to-end distance is equal to $r(\phi, L)$.

Using the fitting parameters in Table 1 we have $q^2/4p < 0.145$ such that $0.86 < \exp(-q^2/4p) < 1$. By approximating this value with unity, we obtain the approximated form of Eq. (16) as

$$Y \approx \frac{L-r}{L-r(\phi,L)} \qquad (17)$$

At initial time ($r = L$) we have $Y = 0$ and at $t \to \infty$ ($r = r(\phi, L)$) we obtained $Y = 1$.

We also conducted a simple simulation to visualize the conformation of the thread at different time. We divided the threads into 1000 identical segments. We place the segments sequentially, the direction of which is selected randomly. We first place the first segment directed horizontally. We also assume the angle made by the i-th and the (i+1)-th segment is assumed to be constant, $\beta$. After placing the first segment, we selected the direction of the second segment randomly by generating a random number $0 \leq w \leq 1$. Suppose the angle made by the i-th segment is $\theta_i$. The angle made by the (i+i)-th segment satisfies the following rule

$$\theta_{i+1} = \begin{cases} \theta_i - \beta & \text{if } 0 \leq w < 0.5 \\ \theta_i + \beta & \text{if } 0.5 \leq w \leq 1 \end{cases} \qquad (18)$$

Inset in Fig. 7(A) is the examples of conformation of the threads when $r = 0.635L$, $r = 0.484L$, and $r = 0.274L$. In simulation we used $\beta = 7°$. It is clearly shown that the conformations produced by simulation nearly match the picture in Figs. 1 and 2.

As final notes we state here that simple phenomena of free falling of threads were able to replicate two famous physical phenomena: two-dimensional random walk and crystallization in one-dimensional space described by the modified Avrami equation. The two later phenomena occur in microscopic scale. Therefore, the present experiment were able to manisfest or replicate two microscopic processes in macroscopic scale that easily observed with naked eyes.

## CONCLUSION

The experiment and model of free falling of threads has been success to bring two microscopic processes (two dimensional random walk and phase transition in one dimensional space according to the modified Avrami equation) into macroscopic scale that easily observed with naked eyes. The dependence of thread end to end distance on the thread length satisfied the scaling relation that precisely similar to the scaling relation in two-dimensional random walk. The equation describing the evolution of thread end to end distance with time is precisely the same as the evolution of crystallization process described by the modified Avrami equation. This is very surprising since a simple phenomenon in our daily life indeed contains very rich physical ingredients.

## ACKNOWLEDGEMENT

The PMDSU (Program Magister Doktor Sarjana Unggul) research grant from the Ministry of Research and Higher Education, Republic of Indonesia No. 1371b/I1.C01.2/KU/2017 for HDR is gratefully acknowledged.